\documentclass[floats,floatfix,showpacs,amssymb,prd,superscriptaddress,twocolumn,aps]{revtex4-1}
\usepackage{graphicx, epsfig, amssymb} 
\usepackage{amsmath, amsfonts}
\usepackage{bm} 

\usepackage[linktocpage]{hyperref}
\usepackage[caption=false]{subfig}
\usepackage[usenames]{color}

\def\be{\begin{equation}}
\def\ee{\end{equation}}
\def\beq{\begin{eqnarray}}
\def\eeq{\end{eqnarray}}

\def\L{\mathcal{L}}

\newcommand{\bea}{\begin{eqnarray}}
\newcommand{\eea}{\end{eqnarray}}
\newcommand{\ben}{\begin{enumerate}}
\newcommand{\een}{\end{enumerate}}
\newcommand{\bi}{\begin{itemize}}
\newcommand{\ei}{\end{itemize}}

\begin{document}

\title{\large Kerr black holes with scalar hair}

 \author{Carlos A. R. Herdeiro}
 \affiliation{Departamento de F\'isica da Universidade de Aveiro and I3N, 
 Campus de Santiago, 3810-183 Aveiro, Portugal.}

 \author{Eugen Radu}
 \affiliation{Departamento de F\'isica da Universidade de Aveiro and I3N, 
 Campus de Santiago, 3810-183 Aveiro, Portugal.}

\date{March 2014} 

\begin{abstract}
We present a family of solutions of Einstein's gravity minimally coupled to a complex, massive scalar field, describing 
asymptotically flat, spinning black holes with scalar hair and a regular horizon. These hairy black holes (HBHs) are supported by rotation and have no static limit. Besides mass $M$ and angular momentum $J$, they  carry a conserved, continuous Noether charge $Q$ measuring the scalar hair. HBHs branch off from the Kerr metric  at the threshold of the superradiant instability and reduce to spinning boson stars in the limit of vanishing horizon area. They overlap with Kerr black holes for a set of $(M,J)$ values. A single Killing vector field preserves the solutions, tangent to the null geodesic generators of the event horizon.  HBHs can exhibit sharp physical differences when compared to the Kerr solution, such as $J/M^2>1$, quadrupole moment larger than $J^2/M$  and larger orbital angular velocity at the innermost stable circular orbit. Families of HBHs connected to the Kerr geometry should exist in scalar (and other) models with more general self interactions. 
\end{abstract}


\maketitle

\date{today}
\noindent{\bf{\em Introduction.}}
Black holes (BHs) are believed to play a central role in many astrophysical processes, ranging from the evolution of stars and galaxies, to powering Active Galactic Nuclei and extreme bursts of gravitational radiation in binary BH coalescence. Central to our understanding of BH physics are the uniqueness theorems~\cite{Robinsoon:2004zz}, stating that the only stationary, regular, asymptotically flat BH solution of the vacuum Einstein gravity is the Kerr metric \cite{Kerr:1963ud}, suggesting that the myriad of BHs in the Cosmos near equilibrium are all well described by this elegant geometry. Such paradigm was summarized by J. Wheeler's ``mantra'': \textit{`BHs have no-hair'}~\cite{Ruffini:1971}.

In this letter we will show that this simple picture has to be reconsidered. We show that a matter field may endow the Kerr metric with `hair', \textit{i.e.} a permanent deformation that keeps its horizon regular and space-time asymptotically flat. The matter we shall consider is a complex scalar field, minimally coupled to gravity and with a mass term. But similar hairy black holes (HBHs) will exist in other scalar field models with more general self interactions. Such scalar fields are ubiquitous in theoretical (astro)physics and may represent a fundamental field or a coarse-graining of fundamental degrees of freedom. As such, scalar HBHs  provide more plausible astrophysical candidates than other known examples of hairy BHs~\footnote{Asymptotically flat, regular BH solutions with hair have been found in Einstein-Yang-Mills theory~\cite{Volkov:1998cc} and in other models, typically with non-linear sources~\cite{Bizon:1994dh,Chrusciel:2012jk}. It is unclear, however, if any of these solutions can have astrophysical relevance.}. 

The HBHs metric presented here remains stationary and axially symmetric, in the same sense as Kerr, but the full solutions - including the matter field - are not preserved by these isometries, being only invariant under the action of a single Killing vector field, which is tangent to the null geodesic generators of the horizon. This is why these HBH solutions are outside the scope of the no-hair theorems \cite{Bekenstein:1996pn}, which apply to stationary solutions. Since the metric of HBHs is stationary, however, they are equilibrium states and may play a role in realistic astrophysical processes.

\noindent{\bf{\em The model.}}
%
The action for Einstein's gravity minimally coupled to a complex massive scalar field $\Psi$ is
$S= \int  d^4x\sqrt{-g}\left[\frac{1 }{16\pi G}R
   - \Psi_{, \, a}^* \Psi^{, \, a} - \mu^2 \Psi^*\Psi
 \right]$,
where $G$, that will be set to unity,  is Newton's constant and $\mu$ is the scalar field mass.
The resulting field equations are the Einstein--Klein-Gordon (EKG) system:
\begin{equation}
\label{EKG-eqs}
 R_{ab}-\frac{1}{2}g_{ab}R=8 \pi T_{ab}, \ (a) \qquad
 \Box \Psi=\mu^2\Psi, \ (b)
\end{equation}
where $
T_{ab}= 
2 \Psi_{ , (a}^*\Psi_{,b)}-g_{ab}  [  \Psi_{,c}^*\Psi^{,c}+\mu^2 \Psi^*\Psi]
$
is the
 stress-energy tensor  of the scalar field. Hitherto, the only BH solution known within this theory is the Kerr family together with $\Psi=0$.

\noindent{\bf{\em Linearized analysis: scalar clouds.}}
 The branching off of the Kerr family into a new family of HBH solutions may be anticipated from an analysis of linearized scalar field perturbations around the Kerr metric. 

Consider the Klein-Gordon equation (\ref{EKG-eqs}b) in the background of the Kerr solution with mass $M$ and angular momentum $J\equiv a M$. In Boyer-Lindquist (BL) coordinates $(t,r,\theta,\varphi)$ variables are separated as
 $\Psi=  e^{-i w t} e^{im\varphi} S_{\ell m} (\theta)R_{\ell m} (r)$, 
 where $S_{\ell m}$ are the spheroidal harmonics, $-\ell\leq m\leq \ell$ and $R_{\ell m}$
satisfies a radial Teukolsky equation \cite{Brill:1972xj,Teukolsky:1972my,Teukolsky:1973ha,Press:1973zz} (`prime' denotes radial derivative)
\begin{eqnarray} 
\left(\Delta R_{\ell m}'\right)'=\left ( a^2 w^2-2ma w+\mu^2 r^2+A_{\ell m} -\frac{K^2}{\Delta}
\right ) R_{\ell m}~. \ \ \ \ \ \nonumber
\end{eqnarray}
Here, $\Delta\equiv r^2-2M r+a^2$, $K\equiv (r^2+a^2)w-am$, 
and $A_{\ell m}$ is a separation constant. We are interested in solutions of the Teukolsky equation that decay at spatial infinity. Since the boundary condition at the event horizon, located at $r=r_H$, is that there is a purely ingoing wave (in a co-rotating frame), BHs do not admit, generically, bound states with real frequency $w$. But they do admit \textit{quasi}-bound states, which have complex $w$ with a negative imaginary part, manifesting that the field is decaying, infalling into the BH.  For the Kerr BH, however, there is a critical frequency:
\begin{equation}
w_c\equiv m\Omega_H \ ,
\end{equation}
defined by the horizon angular velocity $\Omega_H$, such that for $w<w_c$, the imaginary part of $w$ becomes positive. This is the \textit{superradiant} regime~\cite{Press:1972zz,Damour:1976kh,Zouros:1979iw,Detweiler:1980uk,Cardoso:2013krh,Shlapentokh-Rothman:2013ysa,Dolan:2012yt}; the corresponding modes increase in time, signaling an instability of the Kerr BH in the presence of a massive scalar field. 

Precisely at $w=w_c$ the imaginary part of the frequency vanishes and one may expect bound states to exist. Such states, dubbed as scalar \textit{clouds}, were found analytically for extremal Kerr BHs ($a=M$)~\cite{Hod:2012px} (see also~\cite{Hod:2013zza}). They form a discrete set labelled by 3 `quantum' numbers, $(n,\ell,m)$, which are subjected to a `quantization' condition, involving the BH mass. The label $n$ is a non-negative integer, corresponding to the node number of $R_{\ell m}$.  Fixing $(n,\ell,m)$, the quantization condition will yield one (physical) possible value of the BH mass. 

\begin{figure}[b!]
\centering
\includegraphics[height=2.48in]{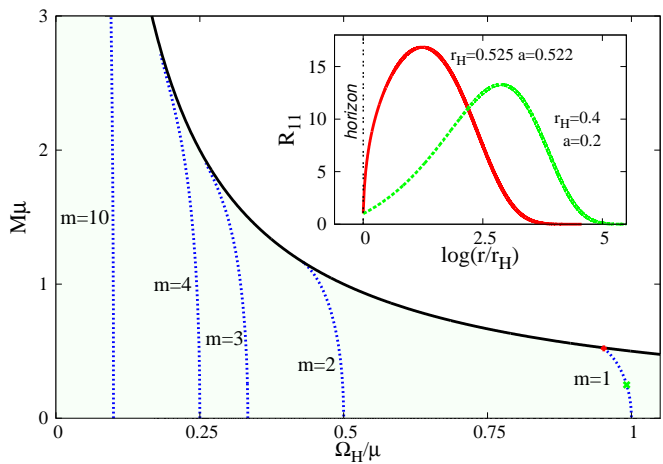}\\
\caption{$M$ vs $\Omega_H$ for Kerr BHs. The black solid curve corresponds to extremal BHs, which obey $M=\frac{1}{2\Omega_H}$; Kerr BHs exist below it (shaded region). A scalar cloud with parameters $(n,\ell,m)$ exists along a line. Five such (dotted blue) lines are shown, for $n=0$, $\ell=m$ and different $m$'s. (Inset) $R_{11}(r)$, from $r=r_H$, and normalized such that $R_{11}(r_H)=1$ for two clouds with $m=1$. The corresponding points in the $m=1$ existence (blue) line are shown with the same colors.} 
\label{linear}
\end{figure}

A similar structure of scalar clouds exists in the general Kerr case, $0<a<M$, as we have checked.  Fixing $\ell,m$, and for a given $r_H$, solutions with proper scalar field asymptotics are found only for a discrete set of values of $a$, each value corresponding to given number of nodes $n$.
Then, fixing $\ell,m$ and following a scalar cloud with a given $n$, defines an \textit{existence line} in the $(M,J)$ space for Kerr BHs. In Fig. \ref{linear} we display such lines in a $M$ versus $\Omega_H$ diagram, for fundamental modes ($n=0$) with $\ell=m$ and several values of $m$. 
Observe that since scalar clouds lie precisely at the threshold of the superradiant instability, the region  to the right (left) of a given line contains unstable (stable) Kerr solutions against the corresponding mode. Additionally, existence lines for modes with a given $m$ and $\ell>m$ (and $n=0$) always lie to the right of the line for $m=\ell$. Thus, the line with $m=\ell$ defines the threshold of the instability for a given $m$. Next, we show that precisely these  $m=\ell$ clouds can be promoted to Kerr `hair'  in the full EKG system.

\noindent{\bf{\em The non-linear setup.}}
To investigate the existence of exact solutions corresponding to deformations of the Kerr geometry by backreacting clouds, we consider a metric ansatz
with two Killing vectors
  $\xi=\partial_t$ and $\eta=\partial_\varphi$,
\begin{eqnarray}
\label{ansatz}
ds^2 &=&e^{2F_1}\left(\frac{dr^2}{N }+r^2 d\theta^2\right)+e^{2F_2}r^2 \sin^2\theta (d\varphi-W dt)^2
\nonumber \\
&&-e^{2F_0} N dt^2, \qquad
~~{\rm with}\qquad ~~N=1-\frac{r_H}{r},
\end{eqnarray}
  where $F_i,W$, $i=0,1,2$, 
  are functions of $r$ and $\theta$ only \footnote{The Kerr solution can also be written in these (non-BL) coordinates.}. The isometry generators  $\xi$ and $\eta$ are not, however, symmetries of the full solution, since we take the scalar field ansatz to be 
\begin{eqnarray}
\Psi=\phi(r,\theta)e^{i(m\varphi-w t)},
\label{ansatz2}
\end{eqnarray}
where $\phi$ is a real function, $w>0$ is the frequency and $m=\pm 1,\pm 2$\dots
is the azimuthal winding number. The fact that the $t,\varphi$ dependence of $\Psi$ occurs as phase factors implies $T_{ab}$ is $t,\varphi$ independent, which is required for the geometry to be stationary and axi-symmetric. $T_{ab}$ will, however, depend on $w,m$ and so will the geometry.

The boundary conditions for the problem are as follows. Firstly, asymptotic  flatness requires that $F_i=W=\phi=0$, as $r\rightarrow \infty$. Then,  the ADM mass $M$ and angular momentum $J$ are read off from the expansion:
\begin{eqnarray}
\label{asym}
g_{tt} =-1+\frac{2M}{r}+\dots,~~g_{\varphi t}=-\frac{2J}{r}\sin^2\theta+\dots. \ \ \ 
\end{eqnarray}
For the scalar field, the asymptotic behaviour must agree with linear analysis:
$\phi=f(\theta)  {e^{-\sqrt{\mu^2-w^2}r}}/{r}+\dots$. Thus, bound states require $w <\mu $.
Secondly, axial symmetry and regularity impose that on the symmetry axis ($\theta=0,\pi$),
$\partial_\theta F_i = \partial_\theta W = \phi = 0$, and, to avoid conical singularities, $F_1=F_2$. 
Finally, near the event horizon, located at $r=r_H>0$, it proves useful to introduce a new
radial coordinate $x=\sqrt{r^2-r_H^2}$.
Then a power series expansion near the horizon 
yields  
$F_i=F_i^{(0)}(\theta)+x^2 F_i^{(2)}(\theta)+{\cal O}(x^4)$,
$W=\Omega_H +{\cal O}(x^2)$,
and $\phi=\phi_0(\theta)+{\cal O}(x^2)$,
with $w=m\Omega_H$, where we take $\Omega_H>0$. 
Thus the boundary conditions at the horizon are  
$\partial_x F_i = \partial_x \phi =  0$ and
$W=w/m$.
Observe that the Killing vector $\chi =\xi+\Omega_H \eta$ is null on the horizon 
and that the  condition $w=m\Omega_H$
implies that there is no scalar field flux
into the BH, $\chi^{\mu}\partial_\mu \Psi=0$. 

The field equations \eqref{EKG-eqs} with the ansatz \eqref{ansatz}-\eqref{ansatz2}, yield a set of five second order,  coupled, non-linear partial differential equations for the functions $F_i,W$ and $\phi$. We have solved them  numerically, subject to the above boundary conditions. We shall focus on even solutions under $\theta\rightarrow \pi-\theta$. Quantities are presented in terms of dimensionless variables, using
natural units set by $\mu$ (recall $G=1$). As such, the numerical construction of the solutions relies on 4 input parameters:  the winding number $m \geq 1$,
the horizon radius $r_H$, the field frequency $w$ and the scalar field node number $n$. The first and third determine the horizon angular velocity as $\Omega_H=w/m$. We shall only consider nodeless solutions since these are typically the most stable ones.
%

\noindent{\bf{\em Physical relations and checks.}}
We have used the Smarr relation and the first law of BH thermodynamics to test the accuracy of the results; so let us consider these relations. 
The BH horizon introduces a temperature~\footnote{
Note that the Einstein-Klein-Gordon equation $R_r^\theta=8 \pi T_r^\theta$ implies that the difference $F_0-F_1$ is constant on the horizon. The $\theta$--independence of this difference actually works as another test of the numerics.} $T_H=\frac{1}{4\pi r_H}e^{F_0^{(0)}(\theta)-F_1^{(0)}(\theta)}$, and an entropy $S=A_H/4$, with 
$A_H=2\pi r_H^2 \int_0^\pi d\theta \sin \theta  e^{F_1^{(0)}(\theta)+F_2^{(0)}(\theta)}$. 
On the other hand, the scalar field  has a global U(1) symmetry which leads to a
conserved current $j^a=-i (\Psi^* \partial^a \Psi-\Psi \partial^a \Psi^*)$. 
Thus the solutions carry also a conserved Noether charge
\begin{eqnarray}
\label{Q}
Q=\int_{\Sigma}dr d\theta d\varphi ~j^t \sqrt{-g}.
\end{eqnarray}
The ADM quantities $M,J$ are related with $T_H,S,Q$ and with $M^\Psi\equiv - \int_{\Sigma} dr d\theta d\varphi(2T_t^t-T_a^a) \sqrt{-g}$, the scalar field energy outside the BH, through a Smarr formula 
\begin{eqnarray}
\label{smarr}
M=2 T_H S +2\Omega_H (J-m Q)+ M^\Psi . 
\end{eqnarray}
The variation of $M$ can be expressed by the first law:
\begin{eqnarray}
\label{fl}
dM=T_H dS +\Omega_H dJ .
\end{eqnarray}
Based on testing \eqref{smarr}, \eqref{fl} as well as standard convergence tests, we estimate a typical relative error $<10^{-3}$ for the solutions reported herein. These solutions have a regular Kretschmann scalar on and outside the event horizon, whose spatial sections have a squashed sphere geometry, just as for Kerr. 
Further details on numerics will be presented elsewhere \cite{Herdeiro:2014}.

\noindent{\bf{\em The HBHs phase space.}}
Solutions are found, after fixing $m$, by sweeping the $w,r_H$ space. For the same value of $w$ there is, in general, an interval of values of $r_H$ for which solutions are found. Then, for each solution, $M,J,Q$ can be computed from \eqref{asym}, \eqref{Q}. In the following we shall use $q\equiv mQ/J$ to parameterize the space of solutions, since it has a compact domain $q\in [0,1]$; $q=0$ is the Kerr limit and $q=1$ is the horizonless solitonic limit -- \textit{boson stars} (see~\cite{Schunck:2003kk,Liebling:2012fv} for a review), which obey $J=mQ$ \cite{Schunck:1996he,Yoshida:1997qf,Kleihaus:2005me}.

\begin{figure}[h!]
\centering
\includegraphics[height=2.48in]{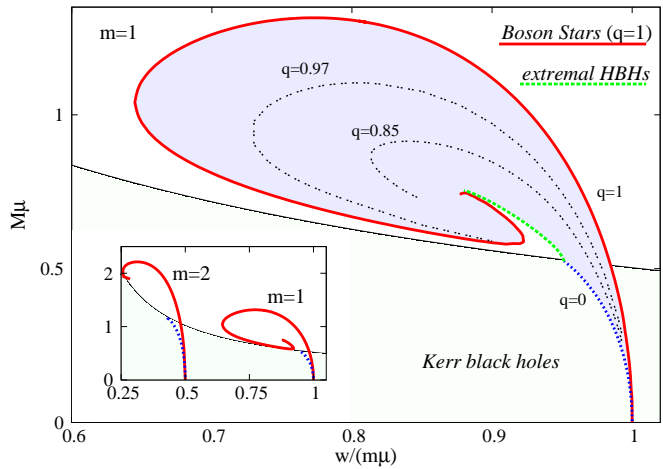}\\
\caption{Domain of existence of HBHs for $m=1$ in $M$-$w$ space (shaded blue region). The black solid line and the dotted blue lines are the same as in Fig. \ref{linear}. (Inset) The boson star lines for $m=1,2$.} 
\label{non-linear}
\end{figure}

For a direct comparison with Fig.~ \ref{linear}, we plot the phase space of HBH solutions in a $M$ versus $w/m=\Omega_H$ diagram -- Fig.~\ref{non-linear} -- obtained from several thousands of solution points. The large plot focuses on $m=1$ solutions; their domain of existence is the light blue region. For $q=0$, it precisely connects to Kerr solutions that admit scalar clouds with $m=\ell=1$ (dotted blue line, c.f. Fig.~\ref{linear}). This shows that the HBHs are indeed the non-linear realization of the scalar clouds obtained in linear theory. For $q=1$, on the other hand, $r_H$ vanishes and HBHs reduce to boson stars (red solid line). The boson star curve in this diagram spirals inwards into a small central region where numerics become increasingly challenging.  The final line that delimits the domain of existence of HBHs (dashed green line) corresponds to extremal HBHs ($T_H=0$). The same pattern occurs for other values of $m$. The inset in Fig.~\ref{non-linear} shows also the existence lines and boson star curves for $m=1,2$.

\begin{figure}[b!]
\centering
\includegraphics[height=2.48in]{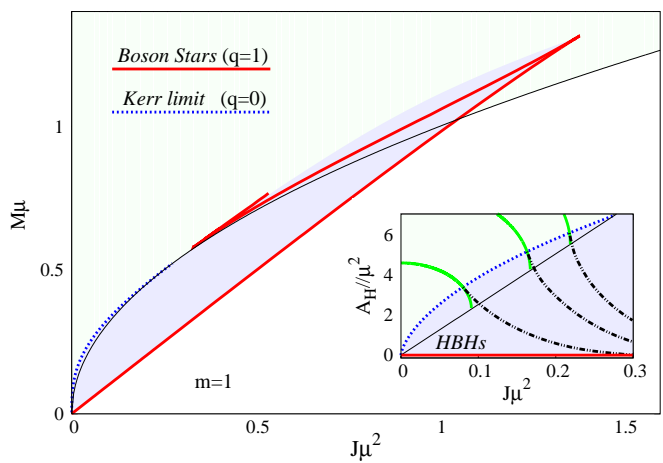}\\
\caption{Domain of existence of HBHs for $m=1$ in $M$-$J$ space. The color code is the same as in Figs.~\ref{linear}-\ref{non-linear}. (Inset) Area as a function of $J$ along constant $M$ curves. Solid green (dashed black) curves correspond to Kerr (HBHs). For the same $M$ they bifurcate in the (dotted blue) Kerr line.} 
\label{non-linear_full2}
\end{figure}

The domain of existence of $m=1$ HBHs in the $M$-$J$ space is exhibited in Fig.~\ref{non-linear_full2}. 
From this figure it becomes clear that there are HBHs with the same $M,J$ as Kerr solutions. The latter exist above the black solid line, which, as before, corresponds to extremal Kerr. In this sense, and because $M,J$ are the only global charges, there is \textit{non-uniqueness}. Moreover, as exhibited in the inset, in the region of non-uniqueness HBH can have the largest area/entropy for the same $M,J$, showing that HBHs cannot decay adiabatically to Kerr. Our numerical analysis indicates, however, that further specifying $q$ raises this degeneracy: no two distinct solutions with the same $(M,J,q)$ could be found. Another observation from Fig.~\ref{non-linear_full2} is that HBHs can violate the Kerr bound $J\le M^2$. This is not surprising. Indeed, boson stars can violate this bound~\cite{Ryan:1996nk}; since HBHs are continuously connected to boson stars, the same should occur for at least some HBH solutions with $q$ close to unity. Fig.~\ref{non-linear_full2} confirms this expectation.

\begin{figure}[h!]
\centering
\includegraphics[height=2.48in]{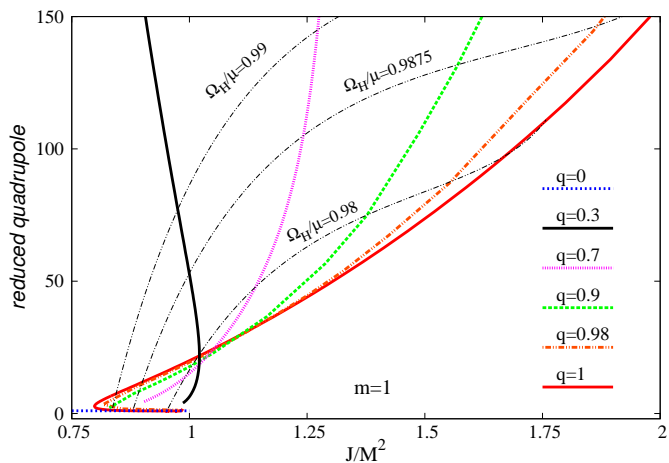}\\
\caption{ Ratio of the quadrupole moment of HBHs to that of a Kerr solution with the same $M,J$. Several lines of constant $\Omega_H$ (dashed black) and $q$ (c.f. caption) are displayed.}
\label{quadrupole}
\end{figure}

\begin{figure}[h!]
\centering
\includegraphics[height=2.48in]{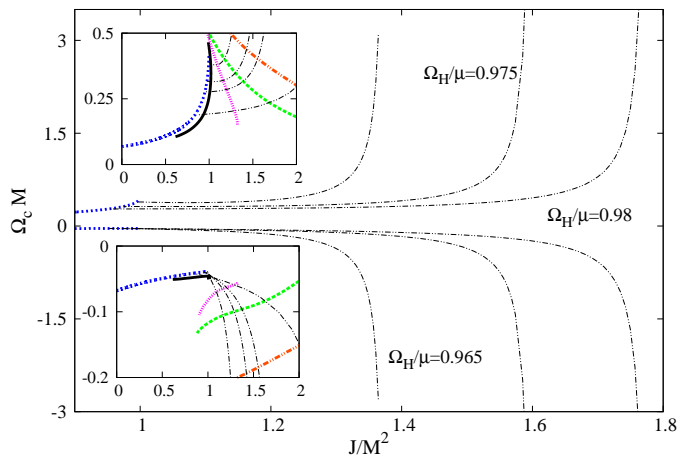}\\
\caption{ Orbital frequency for counter- (bottom curves) and co-rotating (top curves) orbits at the ISCO for Kerr ($q=0$) and HBHs ($q>0$). The color code is the same as in Fig.~\ref{quadrupole}.} 
\label{isco}
\end{figure}

\noindent{\bf{\em Further physical aspects of HBHs.}}
The fact that they are continuously connected to boson stars implies that HBHs are more \textit{star-like} than Kerr BHs; \textit{i.e.} their physical properties are less constrained. Starting from $q=1$ solutions, and slightly decreasing $q$, one is effectively placing a small rotating BH inside a boson star. Thus, the physical properties of the spacetime do not change dramatically. For instance, boson stars can have  quadrupole moments hundreds of times larger than the Kerr geometry~\cite{Ryan:1996nk}, for which the quadrupole moment is determined by $M,J$ as $-J^2/M$~\cite{Hansen:1974zz}. We find an analogous situation for HBHs, as illustrated in Fig.~\ref{quadrupole}.
Here, the quadrupole moment is the Geroch-Hansen~\cite{Geroch:1970cd,Hansen:1974zz} quadrupole, computed with the method of \cite{Pappas:2012ns}. In general, in a stationary axi-symmetric geometry, the multipolar moments determine the observable properties (frequency and number of cycles) of the gravitational radiation emitted by a slowly inspiralling small mass object. As such, large quadrupole deviations relatively to Kerr leave observational signatures in gravitational wave physics.
Another spacetime property that has impact on a number of astrophysical observations is the orbital frequency  at the innermost stable circular orbit (ISCO). For Kerr, in BL coordinates, the ISCO varies from $r=6M$ to $r=M$ ($9M$) for co-rotating (counter-rotating) orbits~\cite{Bardeen:1972fi}, as $J$ grows from $0$ to $M^2$. Consequently, the orbital frequency, $\Omega_c=d\varphi/dt$, at the ISCO varies with $J$. In Fig.~\ref{isco} we plot the corresponding results for HBHs. 
Significant deviations of $\Omega_c$ from the Kerr values leave imprints in astrophysical observations; for instance, in synchrotron radiation from accretion disks. 
%

\noindent{\bf{\em Further implications.}}
The potential (astro)physical relevance of HBHs will depend on a number of factors. 
One such factor is their mass. As it is manifest from Fig.~\ref{non-linear} these solutions have a maximal mass $\sim M_{Pl}^2/\mu$, which will only reach the solar mass if $\mu\lesssim 10^{-11}$ eV. Particles within this mass range are not known, but have been suggested within the String Axiverse~\cite{Arvanitaki:2009fg}. Scalar field models with more general interactions can, however, alleviate this requirement on $\mu$~\cite{Liebling:2012fv}. A second factor is their stability. In general, HBHs have an ergoregion. Does this imply the existence of superradiant instabilities? What are the corresponding time scales? Answering these questions will be crucial to understand HBHs.


%

HBHs connect Kerr BHs to boson stars. As such, they may be presented in two complementary perspectives. Firstly, from the boson stars perspective, HBHs clarify a lingering issue: that a small BH can be added at the centre of a boson star, as it can for other solitons. The crucial new ingredient, in this case, is that the solution \textit{must be spinning}.  Secondly, from the BH perspective, HBHs exist because of the superradiant instability of Kerr, which justifies why HBHs have no static limit. The branching off of a family of solutions of Einstein's equations into a new family of solutions at the onset of an instability is a recurrent situation.  An earlier example is the Gregory-Laflamme instability~\cite{Gregory:1993vy} of black strings, which branch off to a family of non-uniform string solutions at the onset of the instability~\cite{Gubser:2001ac,Wiseman:2002zc}. This perspective suggest that any field (and not just scalar fields)  triggering a superradiant instability will induce BH hair. A non-scalar field example is the Proca field~\cite{Pani:2012vp}; thus BH solutions with Proca hair should exist. Other generalizations of HBHs include gauging the scalar field to obtain hairy Kerr-Newman BHs or imposing $AdS$ asymptotics -- even with $\mu=0$, c.f. the first example in~\cite{Dias:2011at}. Indeed, HBHs substantiate the parallelism between mass and $AdS$ asymptotics as confining mechanisms.

\noindent{\bf{\em Acknowledgements.}} We thank funding from the FCT-IF programme and the grants PTDC/FIS/116625/2010,  NRHEP--295189-FP7-PEOPLE-2011-IRSES.

\bibliographystyle{h-physrev4}
\bibliography{letter}
\end{document}